# Exploring Embodied Emotional Communication: A Human-oriented Review of Mediated Social Touch

Liwen He[1], Zichun Guo[2], Yanru Mo[3], Yve Wen[4], Yun Wang[1*]

[1*]School of Mechanical Engineering and Automation, Beihang University, Beijing, 100191, China.
[2]College of Arts and Design, Beijing University of Chemical Technology, Beijing, 100029, China.
[3]Faculty of Industrial Design Engineering, Delft University of Technology, Delft, 10587, Netherlands.
[4]School of electronic information and electrical engineering, Shanghai Jiao Tong University, Shanghai, 200240, China.

*Corresponding author(s). E-mail(s): wang yun@buaa.edu.cn;
Contributing authors: liwen he@buaa.edu.cn; guozichun@buct.edu.cn; Y.Mo-1@student.tudelft.nl; wenyve17@sjtu.edu.cn;

**Abstract**

This paper offers a structured understanding of mediated social touch (MST) using a human-oriented approach, through an extensive review of literature spanning tactile interfaces, emotional information, mapping mechanisms, and the dynamics of human-human and human-robot interactions. By investigating the existing and exploratory mapping strategies of the 37 selected MST cases, we established the emotional expression space of MSTs that accommodated a diverse spectrum of emotions by integrating the categorical and Valence-Arousal model, showcasing how emotional cues can be translated into tactile signals. Based on the expressive capacity of MSTs, a practical design space was structured encompassing factors such as the body locations, device form, tactile modalities and parameters. We also proposed various design strategies for MSTs including workflow, evaluation methods, and ethical and cultural considerations, as well as several future research directions. MSTs' potential is reflected not only in conveying emotional information but also in fostering empathy, comfort, and connection in both human-human and human-robot interactions. This paper aims to serve as a comprehensive reference for design researchers and practitioners, which helps expand the scope of emotional communication of MSTs, facilitating the exploration of diverse applications of affective haptics, and enhancing the naturalness and sociability of haptic interaction.

**Keywords:** Tactile interface, Affective haptics, Emotional communication, Mediated social touch

# 1 Introduction

Touch is a fundamental aspect of human development. It serves not only as the earliest sensory modality in infancy but also as a significant tool for learning. The perception and comprehension of tactile information profoundly influence human existence and communication in the world [1]. Across various emotional experiences, social touch shapes social rewards, attachment, cognition, communication, and emotional regulation from infancy to adulthood [2]. Neuroscience and social science have extensively explored the relationship between touch and emotions [3]. Tactile interaction is especially crucial in social interactions, where the emotions elicited by physical stimulation can infuse communication with passion, enhance emotional intimacy, closeness, love, vulnerability, a sense of proximity, and the establishment of human connection [4, 5]. Van Erp et al. [6] define this impact of tactile interaction on social interaction as "social touch" and further describe it as the brain's interpretation of pressure, vibration, stretch, and temperature stimulus characteristics.

Tactile technology has the potential to enhance emotional communication in social scenarios. Recent research in the design of tactile technologies has emerged as a novel direction in the field of human-computer interaction. Tactile interfaces have gained rapid momentum and are considered highly promising for compensating for the loss of nonverbal cues inherent in predominantly visual and auditory communication media [7]. While some studies have explored olfactory and gustatory technologies, their applicability in communication remains unclear [8].

The concept of *Affective Haptics* was introduced in 2008, defined as an emerging research area focused on designing devices and systems that could influence human emotional states through the sense of touch [9]. Affective Haptic System Design (AHSD) involving psychological and sociological considerations has been identified as an interdisciplinary domain [10]. Scholars like Jewitt et al. [8] use the term *Digital Touch* to describe tactile interaction technologies within digital media, encompassing interactions beyond the hands. Huisman, Haans, and others have extended the concept of social touch to the realm of tactile technology, coining the term *Mediated Social Touch (MST)* as the ability to convey touch remotely using tactile or kinesthetic feedback technology [7, 11]. Irrespective of the terminology, researchers agree that integrating technology and touch is beneficial for physical and emotional well-being and aids in societal development and communication.

In this paper, we employ the term *Mediated Touch* to distinguish tactile interactions facilitated by relevant technologies from natural touch behaviors. Our focus lies in the domain of MST, specifically exploring the use of tactile technologies for socio-emotional communication [12]. MSTs faces various challenges in emotional communication due to the inherently open and multifaceted nature of tactile channels. Different contexts demand distinct features from tactile interaction systems, designers needs to align with users' interaction intentions and content boundaries to avoid unintended complications. Moreover, understanding the selection of appropriate tactile parameters according to contexts becomes crucial knowledge for designers.

Our work centers on how researchers and designers leverage human tactile perception for emotional communication, with a specific emphasis on the design applications of mediated social touch in conveying emotional information within social settings. Our exploration encompasses the following aspects:

- **RQ1.** What constitutes the emotional expression space of MST within social contexts?
- **RQ2.** How are the design factors of MST interconnected with its emotional expression space?

Taking a human-oriented perspective, this paper systematically organizes previous design strategies employed for affective haptic interactions in MST, with emotion types and body locations serving as key guiding principles. Both the categorical and dimensional models are integrated to elucidate the emotional expression space of existing MSTs, based on which a detailed design space of tactile modalities and parameters is established and various related design strategies are proposed as practical reference guidelines. We aspire to identify potential avenues for future development in this domain and encourage researchers to delve deeper into uncharted territories within MST.

## 1.1 Paper structure

The subsequent sections of this paper unfold as follows (see Figure 1): Section 2 delves into an exposition of the intricate interplay between social contexts, emotional communication, and tactile interaction interfaces. Our approach here is rooted in contextual analysis, setting the stage for our design perspectives. In Section 4, we investigate the expression space of existing MSTs, elucidating applied tactile-emotion mapping strategies, the thematic content of conveyed emotional information, and the types of social relationships that are enhanced by MSTs. Section 5 delves into the expansive design space of MST, with a specific focus on body locations that serve as the nexus for tactile interactions. This exploration encompasses aspects like form factors, tactile modalities, and the finely tuned physical parameters that intricately shape user experiences. Section 6 synthesizes persistent research queries within the domain, spanning suggested design workflow, evaluation methods, as well as the ethical and cultural considerations.

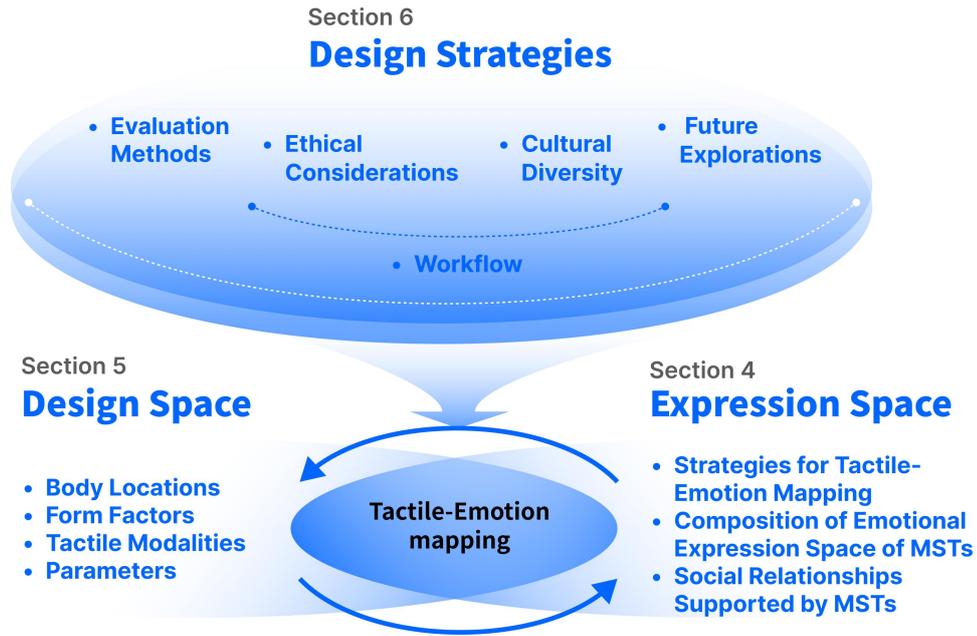

**Fig. 1** The research structure.

## 2 Background

### 2.1 Social interaction via touch

Touch plays a pervasive role in social interactions, and physical sensations contribute significantly to social communication. Through this form of physical interaction based on body locations, individuals express themselves through discrete, brief touches, revealing aspects of themselves such as gratitude, compliance, and other emotional nuances [13–16]. Tactile signals also have a soothing effect within social environments, aiding individuals in adapting to social settings [17].

The significance of touch lies in its ability to evoke emotional responses during social encounters. Bidirectional connections exist between tactile behaviors and emotions [3]: emotions can be conveyed through touch [18], and touch-based interactions can alter emotional states [19]. Tactile stimuli activate the lateral temporal and frontal regions closely associated with emotions [20]. CT fibers, which represent the biological substrate for emotion transmission through touch, enable the transmission of emotion-related activity from the insular cortex to the orbital frontal cortex via tactile channels [21, 22], potentially inducing pleasurable sensations. Humans are capable of discerning common emotions solely through touch [18]. For instance, researchers found that emotions such as anger, fear, disgust, love, gratitude, sympathy, happiness, and sadness can be effectively conveyed through touch applied to the forearm [23]. Furthermore, even when touch is initiated by an unfamiliar individual, recipients can accurately identify various emotions conveyed through tactile interactions [23]. Additionally, researchers have investigated touch's impact within specific contexts; for instance, Bryan et al. developed constructs such as touch self-efficacy (TSE) and touch anxiety (TANX) to examine touch-related individual differences in workplace settings [24]. TSE reflects an individual's self-assessment of their ability to use physical touch to enhance interpersonal communication efficiency in the workplace, while touch anxiety (TANX) is defined as the fear and concern about negative outcomes and discomfort associated with touching others in the workplace. Physiological touch anxiety evaluates an individual's bodily reactions to uncomfortable touch-induced anxiety [24].

Social touch facilitates the development of social relationships. Tactile perception and interaction are core sensory and communicative aspects of the human experience. Specifically, pro-social emotions can transform the indifference of dominant groups into positive actions, contributing to harmonious group dynamics [25]. Touch holds a vital role in forming and maintaining intimate relationships. Physical expressions of affection, especially romantic physical affection (PA), are closely linked to overall relationship satisfaction. For example, actions such as back-rubs, caressing, cuddling, holding hands, hugging, and kissing are significantly related to romantic relationship satisfaction [26]. Touch also serves a regulatory role within work-based relationships. In workplace settings, while touch has been associated with negative effects such as sexual harassment, it also has positive implications, fostering positive organizational relationships [24].

In summary, social interaction through touch is a fundamental aspect of human communication, with tactile signals conveying a wide range of emotions and facilitating social bonding.

## 2.2 Emotional communication via MSTs

Research indicates that technology-mediated touch can evoke emotional responses similar to those experienced through natural touch, thus exhibiting universal haptic empathy [27]. Mediated touch is capable of mimicking physiological responses of natural touch, as demonstrated by Cabibihan et al., who found no significant difference in heart rate variability between simulated and human-initiated touch [28]. Digital touch technologies have the potential to enhance, complement, expand, and even redefine the sense of touch and tactile communication, offering novel avenues for emotional communication and broadening communication modalities. Relevant design research has seen accelerated growth over the past decade, with applications spanning diverse contexts such as health, communication, emotional exchange, and entertainment [8].

In recent years, there has been a rapid development in studies on mediated interaction. Enhanced sensory experiences provided by technology have gained prominence, driven by insights into human sensory cognition and influencing the development of digital technologies [8]. Technology has been employed to overcome distance barriers, facilitating emotional communication pathways with peers. Mediated interaction often involves fewer channels (e.g., text or audio only), resulting in limited exchange of contextual and nonverbal cues. Furthermore, media, especially those with lower richness, often fail to provide users with a sense of social presence [29, 30], making it difficult to effectively convey mutual influence during interpersonal interactions.

The emergence of MST expands the information exchange pathway for traditional perceptual media. Just as affective haptics was first defined by Tsetserukou et al. in 2009 as "the emerging area of research which focuses on the design of devices and systems that can elicit, enhance, or influence on emotional state of a human by means of sense of touch" [9], tactile technology provides new avenues for emotional communication. Mediated touch can enrich experiences and increase the amount of information communicated without conveying redundant information, compensating for losses in nonverbal cues due to the use of current communication media [31]. MST is well-suited for social scenarios due to the presence of various nonverbal and non-visual cues inherent in interpersonal interactions, especially in intimate settings [32]. Furthermore, mediated touch communication is user-to-user, maintaining privacy and suitability for exchanging sensitive information [33], particularly in quiet environments like libraries. Additionally, compared to overloaded channels such as visual and auditory ones, the tactile channel often remains dormant, making it suitable for receiving brief signals and providing timely responses [34]. Surface tactile technologies based on skin touch can be tangible, wearable, and technically mature, offering features like high resolution, rapid responsiveness, and strong information-carrying capacity. As such, it offers a novel means of engaging in social communication [22].

In summary, the current state of emotional communication through MSTs demonstrates a burgeoning interest and investment in technology-facilitated interactions designed to evoke emotional responses resembling those of natural touch experiences. Moving forward, future research must concentrate on enriching the depth and social resonance of mediated touch interactions, ensuring their suitability for refined emotional communication within social contexts.

### 2.3 Typology of MSTs

Affective haptic interaction technologies are designed based on the emotional communication path and tactile interaction loop of humans. These technologies address the functional demands of each segment in the interaction loop. In a natural state, emotional communication tasks through the tactile channel involve two directions: (i) Brain-Encode Emotional Information → Tactile Action-Express Emotional Information and (ii) Tactile Perception-Receive Emotional Information→ Brain-Decode Emotional Information. As an externally constructed interaction interface, MSTs serves as an agent for emotional communication between the two ends of the interaction. In these segments, mediated touch employs (i) a haptic display for Tactile Action-Express Emotional Information, presenting tactile signals encoded by the human brain to actively or passively allow the tactile receiver to obtain emotional information conveyed by the haptic display, and (ii) a haptic sensor for Tactile Perception-Receive Emotional Information, capturing tactile-related behaviors of the human on the other end and processing them to understand the emotional information in the tactile behaviors.

### 2.3.1 Touch sensor: Emotional information acquisition

Detecting emotional states through tactile behaviors is a less intrusive form of affective sensing technology [35]. Exploring the mapping mechanisms between touch behaviors and emotions, and identifying touch behavior features affecting emotional expression, helps in effectively measuring user emotions. Tactile sensing acquires tactile information, including physiological information, tactile behaviors, and gestures.

Researchers in human-computer interaction build systems for dynamic emotion detection. Analyzing user features like facial expressions [36], vocal expressions [37], head movements, body gestures [38], and cursor movements [39, 40], helps differentiate emotions using tactile behaviors, e.g., pressure characteristics [41–43]. Experimental work characterizes emotions via clicking behavior duration and acceleration [41], showing shorter joyful clicks and longer sad ones. Touch behavior detection can extend to smart home environments, using touch sensors to capture emotional states for responsive adjustments, e.g., music player mood recommendations and emotional-aware smart floors.

### 2.3.2 Touch display: Emotional information expression

Human sensory systems support discriminative and affective inputs. The tactile channel plays a role in perceiving emotions. Tactile technology advances, applied in HCI for emotional arousal and implicit emotional interaction [44]. Evoking emotional states via tactile technology enhances natural social tactile design for social robots, aiming for emotional goals [45]. Tactile simulation technology regulates emotions, as seen in gentle dynamic touch's pleasantness [46]. Tactile technologies, like tactile messages and virtual touch [47], enhance communication, intimacy, and security. Social robots, e.g., the "Haptic Creature" [48–50], convey emotional arousal via tactile sensations like weight, warmth, and vibrations, mapped into an emotional expression space.

Vibrators and force feedback devices implement tactile technology. Force feedback devices provide realistic sensations [44]. Vibrators evoke social emotional information [51]. Vibrational tactile stroking parallels natural stroking's pleasure [51]. Tactile technology's emotional evocation factors include speed, intensity, action time, surface roughness, friction, softness, and smoothness [46, 52].

The passive acquisition of human emotional information expands into conveying and eliciting emotions. This influences remote human-to-human and human-to-machine interaction, focusing on tactile display systems in MSTs.

## 2.4 A human-oriented design perspective for MSTs

Numerous prior endeavors have predominantly stemmed from the domain of tactile technologies [53–55] in the pursuit of identifying appropriate scenarios for tactile interactions. This is largely attributed to the fact that researchers in this field typically possess expertise in tactile technology. Additionally, the field of tactile interaction is currently undergoing a technology-driven phase. Consequently, when designing interfaces for tactile interaction, mechanical parameters of tactile technology often serve as initial design considerations. More recently, a growing number of researchers with a human-centric focus have ventured into the realm of tactile interaction design. To gain insight into tactile devices before embarking on user experiments, Huisman [56] conducted a comprehensive review of the technical intricacies and applications of tactile technology in the context of social touch.

Looking ahead, as technology matures, the role of the tactile modality within affective computing is gaining increased attention. Tactile interaction is poised to transition into a phase of user-experience-driven design. Following a design-oriented perspective [57], designers often adopt user-centered design methods [58, 59] commonly practiced in the field of Human-Computer Interaction (HCI), which places application scenarios at the forefront of design considerations. Skin is no longer viewed solely as an organ to uncover its sensory receptors (nerve endings and corpuscles), but reconceptualized as a boundary-laden nexus, intrinsically endowed with socio-cultural attributes [60]. Specifically, designers may integrate tactile interaction into existing products, often predefined with tactile interaction nodes. Alternatively, they may infuse tactile channels to convey emotional interactions in distinct interactive scenarios, thereby predetermining the contours of affective interaction content. Building on a human-oriented perspective, our aim is to systematically organize the literature and provide design insights and practical guidance for human-oriented tactile interaction design.

In summary, we have chosen *Emotion types* and *Body locations*, both closely related to humans, as guiding principles in this paper (see Figure 2), based on the involvement of two crucial components of human interaction: (i) mental (emotion responses): where emotional responses occur during interaction, and (ii) physical (body locations): where the body interacts physically with the machine. Our focus is on deciphering how designers and users employ mediated touch for communication. This entails a comprehensive exploration encompassing digital touch technologies, mediation models, and the nuanced aspects of tactile-emotion mapping design [8].

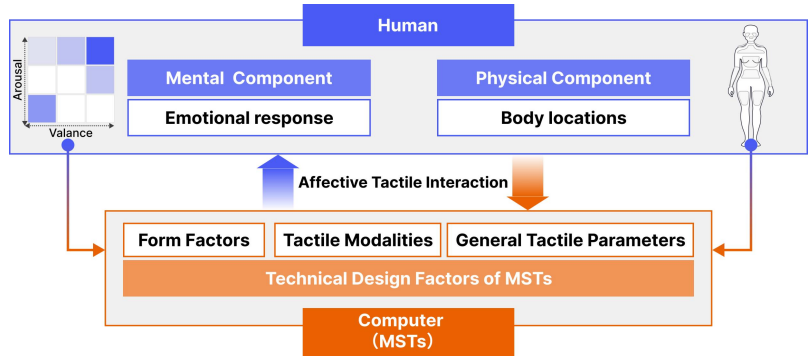

**Fig. 2** Human-computer interaction process via MST and the key factors.

# 3 Methodology

## 3.1 Scope definition and preliminary exploration

To outline our research scope, we initially examined recent review articles from the past five years. Gathering these up-to-date review articles has assisted us in comprehensively understanding the current advancements in tactile interactions. Subsequently, we expanded our search to literature dating from 2000 to the present, utilizing keywords such as "social touch," "digital touch," "affective haptics," and "mediated touch," among others. to conduct preliminary searches across diverse plat- forms, including Google Scholar (comprehensively), ACM Digital Library (primarily in Human-Computer Interaction), IEEExplore (with a focus on haptic technologies), and other pertinent sources. This iterative process refined our keyword repertoire, enhancing its efficacy in guiding subsequent case collection. Prior to embarking on an extensive literature search, we curated a select set of approximately ten mediated social touch studies. This initial curation facilitated discussions on framework organization and enabled us to conduct preliminary analyses.

## 3.2 Data collection

In the emerging interdisciplinary research field of social touch, there is a diverse array of literature types. To ensure comprehensive coverage of various literature types and sources, we employed a combination of methods including keyword searches, manual searches, and citation tracking. Through the initial batch of literature, we refined our keyword set and delineated our screening criteria. Subsequently, employing the refined keyword set, we conducted an expanded search, targeting 145 pertinent literature items spanning from 2010 to 2023. We then proceeded with two rounds of screening (as depicted in Figure 3). In the first round, we selected articles discussing mediated tactile interactions, excluding those focused solely on (i) physiological principles of touch and (ii) natural touch behavior studies. In the second round, we employed modalities of tactile technologies for further screening, excluding articles pertaining to kinesthetic haptic devices and retaining those focused on tactile interaction technologies based on cutaneous haptics. In the third round, we primarily filtered articles based on the application domains of mediated touch systems: articles were retained if they applied to (i) emotional communication scenarios, thereby excluding works solely aimed at enhancing rendering realism without considering emotional effects in tactile displays and those focusing solely on tactile sensor input without analyzing emotional attributes; (ii) interpersonal communication scenarios, thereby excluding works solely aimed at self-feeling reinforcement for media entertainment, human-environment interaction, or human-plants interaction. Additionally, we extended the time frame slightly to include a small number of publications from the period 2000-2010 as supplementary materials.

Finally, 37 MST cases were selected that highly aligned with our research scope, comprising 28 conference papers and 9 journal articles. These cases were primarily sourced from prominent ACM and IEEE conferences (e.g., ACM CHI, IEEE Haptics

Symposium (HAPTICS)), as well as established journals in human-computer interaction and affective computing (e.g., ACM TOCHI, IEEE ToH). Moreover, insights from publications in education, psychology, and medicine provided valuable perspectives on cultural influences, physiological mechanisms, emotion taxonomy, and subjective assessment tools. Notably, our analysis of cultural influences was enriched by an extended examination of local research in China, incorporating approximately ten cases written in Chinese.

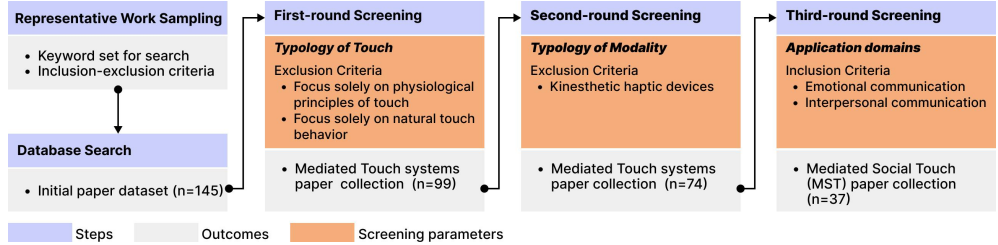

**Fig. 3** Flow chart of literature search and screening.

# 4 Expression space of MSTs

## 4.1 Strategies for tactile-emotion mapping

The design of tactile-emotion mapping mechanisms in MST typically follows two design strategies: (i) existing strategy, which references known natural tactile-emotion response mechanisms, simulating similar emotional meanings by utilizing natural touch behaviors as mediating models; and (ii) exploratory strategy, which explore the expressive capacity of new interaction patterns, defining extended meanings of tactile interaction actions.

### 4.1.1 Existing mapping strategy

54.1% (20 works) of MST designs were guided by existing strategy, referencing physiological signals and social gestures. In these studies, natural touch interactions with emotional significance were enumerated and categorized. Referenced natural interaction behaviors include neutral actions (e.g., Stroke, Squeeze, Poke, etc.), emotional actions (e.g., Caress, Hug, Heartbeat), reactions and sensations (e.g., Goosebumps, Vibrations), and communication and expression gestures (e.g., Ask for attention, Contact) [9, 61–64]. Simulating natural touch through various technological means is explored; for example, force feedback and vibration feedback are used in 11 and 10 works, respectively. These technologies include elastic bands, pneumatic devices, and vibration actuators to simulate behaviors such as hugging, resulting in the pleasure of natural hugs, feelings of social presence, and intimacy [9, 61, 62]. Additionally, vibration actuators are utilized to simulate actions like stroke and squeeze on the forearm or provide fingertip-like click effects on the cheek [63, 64].

### 4.1.2 Exploratory mapping strategy

Exploratory strategy aims to reshape people's understanding of touch and tactile technologies and enhance the application capabilities of modalities like vibration. Among the surveyed works, 45.9% (17 works) adopted this strategy. Vibration feedback is more prevalent when designing innovative mapping relationships with 10 works. This could due to the relatively uncommon tactile effects of vibration actuators in natural touch, making them suitable for novel semantic assignments. Vibrations can produce both categorical and continuous effects, thus enhancing their expressive capacity. For instance, a matrix of vibration actuators generates dynamic, continuous tactile signals on the forearm, belly, or generates sensations akin to "Butterflies in the stomach" associated with love [9, 65]. Similarly, descriptions of awe and fear include references to "Nape's hairs standing on end" [66].

## 4.2 Composition of emotional expression space of MSTs

The expressive capability of social tactile interfaces is reflected in their diversity in conveying emotional information. Researchers build upon limited technological means to express social emotions. Currently, the full extent of the expression space of MSTs remains undetermined. Systems guided by existing mapping strategies can replicate the emotional expression space of natural touch behaviors through explicit validation. However, whether there are emotional effects beyond the natural behavior space and the effectiveness of MST under exploratory mapping strategies requires user verification.

There are over 30 types of emotions expressed through MSTs in our selected studies. Most MST-expressed emotions are categorical, however some works document emotional effects with distributed properties in continuous emotion models. In this paper we integrate both the categorical model and Valence-Arousal model to understand the composition of MST's emotional expression space.

### 4.2.1 Categorical emotions

We compiled the expressed categorical emotions from 37 sources, totaling 36 types (Figure 4). Some terms were grouped due to inconsistent language, but they held consistent emotional meanings, for instance, Joy encompassed Joy/Happiness/Pleasantness/Excitement/Delight, and Content included Contentment/Satisfaction. Among these, positive emotions (41.7%, n=15) and negative emotions (41.7%, n=15) are almost equal, with neutral emotions (16.7%, n=5) being fewer. Frequently mentioned emotions (appearing at least 5 times) include Joy (n=26), Sadness (n=16), Love (n=10), Anger (n=9), Calmness (n=9), Empathy (n=7), Fear (n=6), Gratitude (n=5), Sympathy (n=5), Disgust (n=5). In all conveyed categorical emotions, we selected two subsets for analysis (see Figure 4): (i) the Basic Emotions set: these emotions are general and reflect MSTs' capacity to support emotional communication; (ii) the Prosocial Emotions set: the expression of these emotions facilitates social interaction, and assessing their conveyance reflects MSTs' suitability for social contexts.

Basic emotions play a crucial role in adults' cognitive processes and significantly impact social activities. These emotions interact and interface with higher-order cognitive processes to generate complex emotional experiences and behaviors that are more intricate than primitive emotions alone. Therefore, starting from basic emotions to understand the emotional information set that MST is expected to fulfill is a direct and effective method. Researchers attempt to comprehend emotions at a fundamental level, defining a set of categorical emotions as basic emotions. These emotions exhibit fixed neurological and bodily expressions and are selected based on long-term interactions with ecologically valid stimuli. Notable models include Ekman and Cordaro's [67] *Big Six Emotions* which have clear-cut supporting evidence for each emotion, making them suitable for assessing the basic emotional expression of MST. These include Joy, Sadness, Anger, Fear, Disgust, and Surprise. Basic emotions are mentioned 66 times in the literature (Figure 4), with Joy (n=26), Sadness (n=16), and Anger (n=9) being the most frequently mentioned. The other three basic emotions—Fear (n=6), Disgust (n=5), and Surprise (n=4) are mentioned less frequently.

Prosocial emotions[68] are thought to contribute to prosocial behaviors, such as helping, sharing, and comforting[69]. Further, prosocial emotions can foster social interaction. For instance, emotions like shame, guilt, pride, regret, joy, and other heartfelt reactions show long-term commitment to cooperative outcomes. MST mentions 68 instances of prosocial emotions, encompassing 14 emotions like Love, Empathy, Gratitude, Awe, and Pride. Apart from Joy, Love (n=10), Empathy (n=7), Gratitude (n=5), and Sympathy (n=5) are frequently discussed (Figure 4).

Furthermore, researchers employ MST to convey specific information. For instance, Salvato et al. utilize a vibration actuator matrix in a sleeve to convey the message "Ask for attention", assisting users in perceiving others' appeals in social situations. Critical "Comfort" information can also be conveyed through vibration, pneumatic feedback, and other means, proving invaluable for remote emotional interactions.

### 4.2.2 Integrating categorical emotions within the VA model

Some studies examine the location distribution of emotional responses evoked by social touch in dimensional spaces like the Valence-Arousal (VA) model instead of specific named categories. These continuous models effectively capture the emotional efficacy of social tactile interfaces. The emotional space here is a Cartesian space, where each dimension corresponds to a psychological attribute of the emotion (e.g., the arousal attribute indicating the intensity of the emotion and the valence attribute indicating its positive/negative aspect). Theoretically, this space's descriptive ability encompasses all emotional states, reflecting the strength of emotions' manifestations in each dimension's coordinate values. For example, Feng et al.'s study [70] analyzes emotional responses evoked by their shape-changing interface in a VA model, primarily distributed across four quadrants.

We positioned 36 emotions within the VA model based on the Circumplex model [71] and Plutchik's wheel of emotions [72], examining their distribution. In this process, we simplified the dimensional emotional space by categorizing the arousal axis into Activated, Neutral, and Deactivated, and the valence axis into Pleasant, Neutral, and Unpleasant, following Russell et al.'s emotion labels [71]. Subsequently, these axes intersected to form nine regions. This region division allowed for a basic clustering of discrete emotions to understand emotional distances or identify proximate emotions. The distribution of these emotions in the VA quadrant shows that compared to neutral emotions, strong emotions with a clear tendency for arousal or valence are more expressed in MSTs. For example, emotions with high arousal and valence, such as joy and love, as well as high arousal emotions like anger and low arousal emotions like calmness and sadness. Additionally, high-arousal emotions are mentioned most in MSTs, indicating a greater diversity in the range of high-arousal emotions discussed.

### 4.3 Social relationships supported by MSTs

As a medium for communication in social contexts, the conveyed emotional information through MST is dependent on the emotional communication demands present in different social relationships. In the surveyed works, the majority involve assisting human-human interactions (81.1%, n=30). Early MSTs primarily targeted romantic relationships. With the development of experience design towards interpersonal communication, MST applications expanded to include intergenerational communication. However, MSTs predominantly target interactions between two individuals, with fewer designs addressing multi-person relationships like coworker interactions.

Enhancing partner relationships is a common objective, where simulating touch behaviors such as gripping and stroking enhances the sense of presence[73] and intimacy[74, 75]. In multi-person interactions, MST has the potential to foster perceptions of surrounding individuals in public spaces. For instance, LoveBomb[76] conveys emotions to people within a specific distance range in public spaces, reducing loneliness for emotionally distressed individuals by establishing connections with those nearby. Such tools could positively impact groups like the elderly, prone to loneliness. Additionally, Cang et al.[77] explore the possibility of conveying complex emotional information to strangers using a sleeve. Expanding the target audience of MST poses challenges, the accuracy of recognizing tactile signals may vary based on the intimacy level of relationships, with higher recognition rates for closer partners and lower rates for strangers. However, this assumption varies across different tactile signals and requires further validation.

## 5 Design space of MSTs

### 5.1 Body locations

The selection of interaction locations on user's body requires consideration of various aspects, including (i) the tactile perceptual capabilities of each body location (e.g., the distribution of cutaneous haptics and mechanoreceptors, as well as thermoreceptors. For instance, the spatial resolution of full-body skin tactile perception, as depicted in Figure 6(a), which is determined through two-point discrimination threshold tests), (ii) technical compatibility (e.g., the size and weight of the interaction device), and (iii) social acceptance (e.g., the privacy issues of interaction content) [78]. We compiled data on the selection of tactile interaction locations in the cases (see representative MSTs on various body locations in Figure 5) and systematically analyzed the different emotional communication tasks, enabling designers to find practical references according to their design objectives.

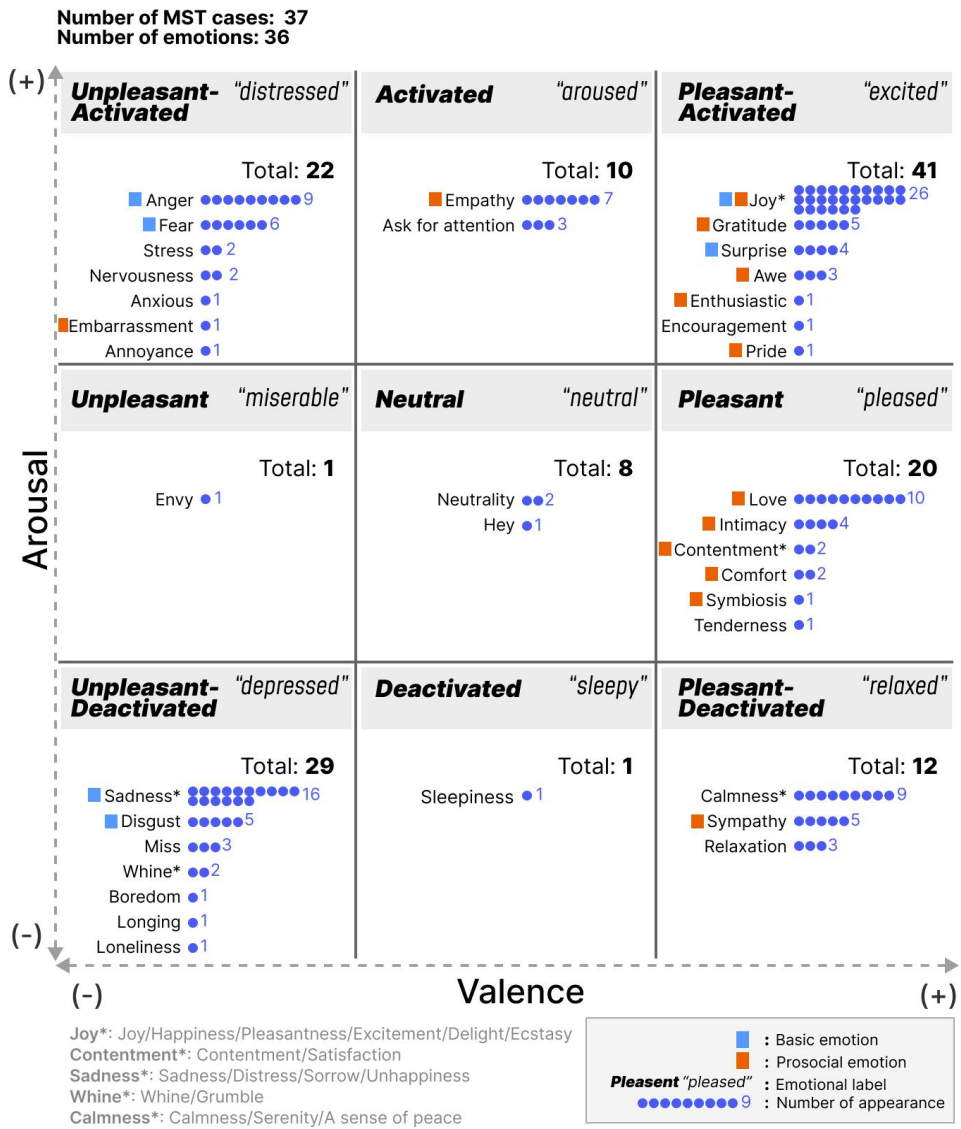

**Fig. 4** The distribution of expressed categorical emotions within the Valence-Arousal (VA) model.

In the 37 surveyed works, MST investigates 14 body areas as receptive locations for emotional tactile interactions, categorized in a hierarchical manner (e.g., cheek, lips, and so forth). In this regard, forearm [12, 63, 65, 77, 79–85], hand [27, 44, 73, 74, 76, 85–91], and back [9, 62, 92–96] have been widely explored with 12, 13, and 7 cases

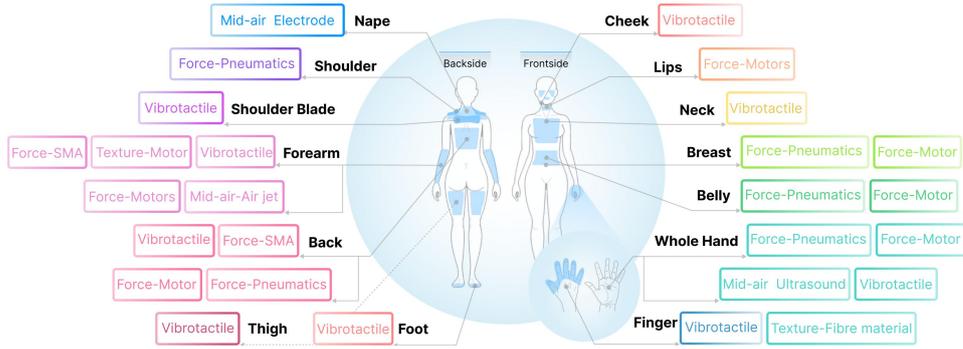

**Fig. 5** The representative MSTs on various body locations.

respectively. Based on interaction characteristics, we classify the 14 body locations into four groups (see Figure 6(a)): (i) *Intimate*: cheek, lips, neck, and nape; (ii) *Immersive*: breast, back, and belly; (iii) *Public*: shoulder, shoulder blade, forearm, entire hand, and fingers; and (iv) *Unusual*: thigh and foot.

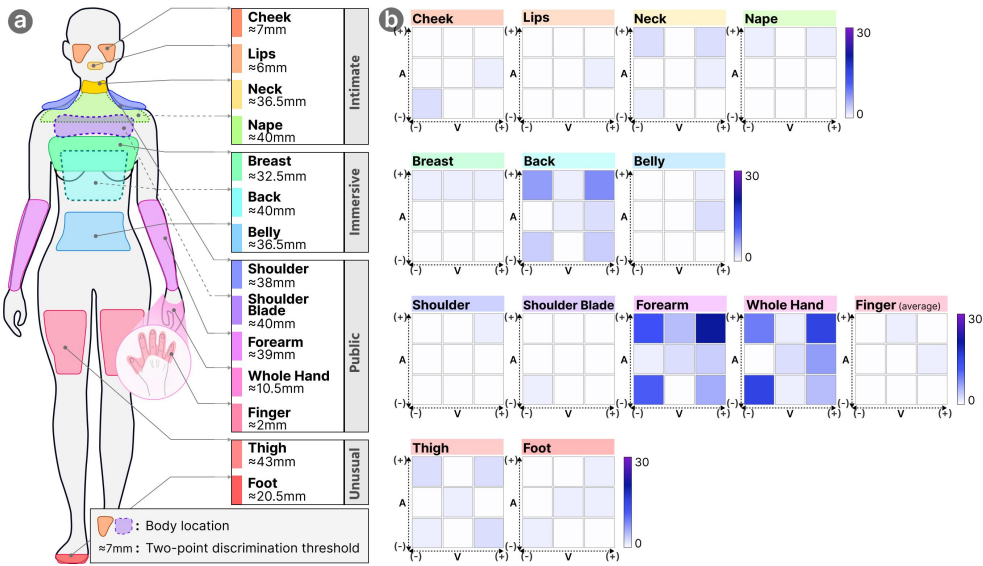

**Fig. 6** (a) Body locations for tactile emotional interaction (including cutaneous two-point discrimination thresholds of each location and the grouping of locations). (b) Emotional information distribution across body locations in the VA model for tactile emotional interaction.

We tabulated the emotional information exchanged at different body locations in the cases, along with their distribution in the VA model. As indicated in Figure 6(b), distinct emotional communication scenarios are facilitated by different body locations with interaction characteristics. For example, tactile interactions among coworkers often occur on areas such as the hands and shoulders, such as shaking hands or patting shoulders. In romantic relationships, interaction commonly involves areas like the cheeks, lips, and torso, such as stroking cheeks, kissing, and embracing. Among these, forearm and shoulder [61, 97] are relatively common areas suitable for both romantic and professional relationships to convey emotional information through touch. Moreover, the high spatial resolution of tactile perception on the forearm allows for the recognition of complex tactile signal patterns, making it a subject of extensive research. Similarly, the hand serves as a primary region for tactile interaction exploration, exhibiting heightened sensitivity to tactile signals and the ability to discern nuanced emotional information, especially within the two-point discrimination threshold of the finger, which is approximately 2 mm. The central trunk area (back) is conducive to providing broad, immersive tactile experiences, such as simulating comforting hugs. It's important to note that although the lips [75], cheek [64], and neck [93] are sensitive to tactile signals, they are generally involved in touch interactions within intimate relationships or specific cultural contexts. Furthermore, the sensitivity of the legs to tactile stimuli is relatively lower. In many cultural contexts, using the foot [98] for interaction might be perceived as impolite. Breast, belly, and other upper-body regions tend to evoke discomfort, offense, and tension when touched, making them less common in social scenarios, and consequently, there are fewer designs targeting these areas.

## 5.2 Form factors

After determining the interaction locations, another significant physical design characteristic is the form factor. We classify MSTs into 4 form categories based on their wearability and mobility: wearable, hand-held (non-wearable, mobile), desktop (non-wearable, immobile), and others (i.e., mobile robots). Each of these interaction device forms possesses distinct attributes.

Among the 37 surveyed works, almost all the body area below the head can adopt wearable form of MST (see Figure 7). Wearable systems hold an advantage for providing tactile feedback as they ensure direct contact with the human skin, mimicking genuine social contact. In contrast, hand-held devices are less common, which could due to the extra effort required from users, such as grasping, leading to unstable contact. Furthermore, researchers have devised "lively" desktop robots capable of conveying emotional information through proactive movement, such as tabletop social robots with multiple texture modules designed by Sato et al. [84], and SwarmHaptics developed by Kim et al. [85], a set of small wheeled robots. On the other hand, non-mobile, fixed desktop tactile displays impose the least physical burden on users and primarily target two areas: forearm [82] and hand [91]. These desktop setups offer mobility convenience and easy posture adjustments to suit tactile interaction. Additionally, for areas like the back and thigh, fixed desktop displays maintain a relatively close proximity to the user during stationary sitting, providing stable opportunities for contact behaviors (e.g., cushion-based interactions [96]).

Regarding the richness of emotional expression, we examined the variety of emotional interaction tasks supported by 4 types of devices (as shown in Figure 7(b)).

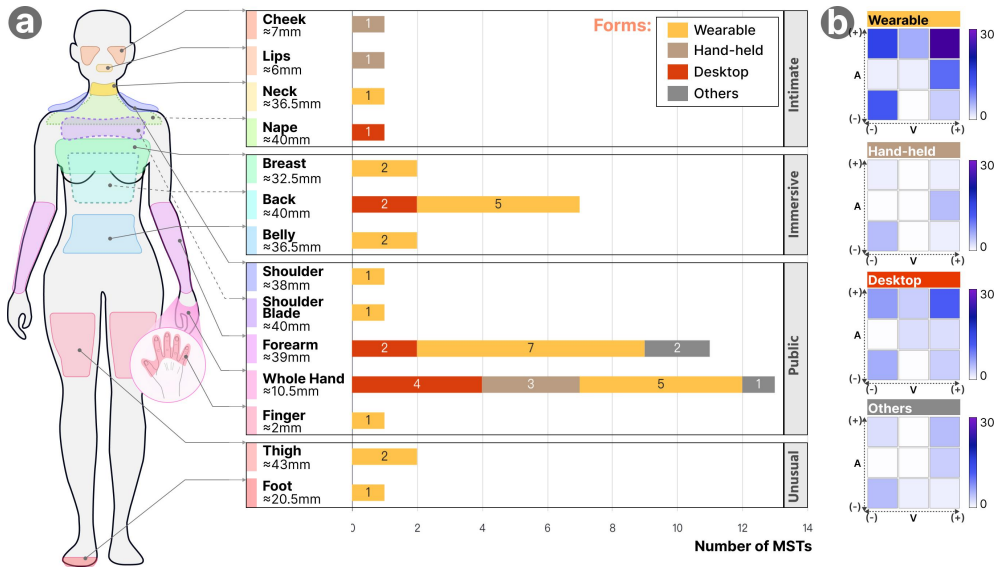

**Fig. 7** (a) The forms of MSTs applied to various body locations. (b) Emotional information distribution across forms in the VA model for tactile emotional interaction.

Wearable devices exhibit the highest richness in emotional expression and have been empirically proven most frequently. Although desktop devices are less numerous in the surveyed cases, they support a diverse range of emotional communication tasks. While designing MSTs, emotions with clearer valance value and higher arousal are often prioritized for expression, whereas much fewer designs are intended to express those with relatively unclear valance and neutral arousal.

### 5.3 Tactile modalities and parameters

The tactile modalities analyzed in this paper include vibration feedback [77, 79], force feedback [82, 88], as well as other tactile feedback forms (such as mid-air modalities [66] and texture-based feedback [90]). Among these, vibration feedback (54.1%, n=20) and various forms of force feedback (43.2%, n=16) are prevalent. Figure 8(a) depicts the application of tactile modalities across different body areas in the literature. Some of these cases are shown in Figure 5.

Similarly, we conducted a statistical analysis of four tactile technology modalities within MST cases to understand their applications in emotional interaction scenarios. As illustrated in Figure 8(b), vibration feedback and force feedback, as the most explored modalities, exhibit the richest emotional expressions and demonstrate a consistent distribution pattern with the overall MST trends: emotions characterized by Pleasant-Activated, Unpleasant-Activated, and Unpleasant-Deactivated are more frequently expressed, with the upper-right quadrant of the two-dimensional VA coordinate space (indicating higher arousal and higher valence) being more prominently activated.

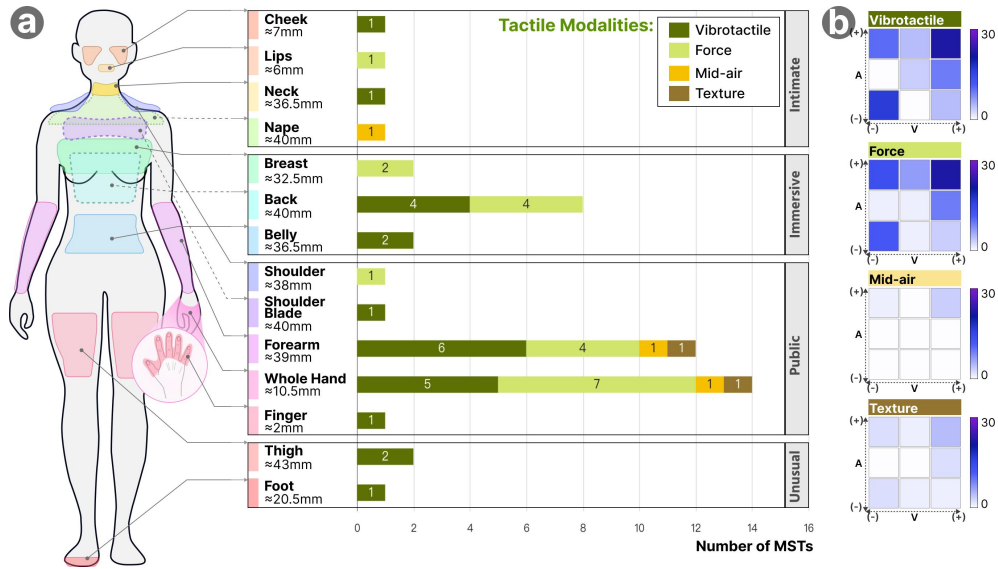

**Fig. 8** (a) The tactile modalities of MSTs applied to various body locations. (b) Emotional information distribution across tactile modalities in the VA model for tactile emotional interaction.

### 5.3.1 Vibration feedback

Utilizing vibration feedback to convey information is a common modality, employing technologies such as Eccentric Rotating Mass Vibration Motors (ERM motors), Linear Resonant Actuators (LRA motors), voice-coil motors, and speakers. Notably, although vibration feedback generates point-like normal effects, researchers have successfully emulated continuous touch actions using discrete vibrational feedback. For instance, Culbertson et al. [65] utilized sequential actions of actuators on the skin to produce a linear movement effect. They found that longer duration of each actuation resulted in shorter delays between adjacent actuators' activation, enhancing continuity.

### 5.3.2 Force feedback

Natural touch actions typically exhibit two main effects: compression and shear, occurring individually or combined, with various parameters such as duration and intensity generating a wide array of emotionally meaningful complex behaviors. Techniques like Pneumatics, motors (employing solid materials or elastic materials such as belts), and Shape Memory Alloys (SMA) are employed to create normal force feedback and tangential force feedback, resulting in compression and shear effects.

Many social tactile interfaces impose forces on users, utilizing passive touch to convey emotional information. This form of tactile information can be conveyed through mechanisms such as (i) rigid levers, as seen in EmotiTactor [82], which simulates arm grasping effects, accurately applying forces in specific locations and supporting both compression and shear effects; (ii) pneumatic devices, such as air jets inducing compression on body parts like hands, arms, and torso, creating a more enveloping normal force feedback that generates dynamic compression [99]; (iii) elastic straps; and (iv) Shape Memory Alloys, generating driving force through deformation.

Moreover, force feedback can also evoke emotional information through active touch perception, for instance, through changes in shape. Thus, various driving techniques can be explored to construct interfaces that manipulate shape for emotional communication [92]. Not limited to static shapes, dynamic parameters of shape alteration can also be incorporated, as demonstrated by Feng et al. [70] who explored the effects of Shape-Change Size and frequency on valence and arousal using the Bouba (rounded)-Kiki (spiky) paradigm.

### 5.3.3 Other modalities

Emerging surface tactile technologies have also been applied in MST. Examples include the use of mid-air modalities such as ultrasound [91] and air-jet technology [83] to induce pleasure in users, and electrodes [66] to evoke emotions like fear and awe. The texture features of materials themselves are also suitable for arousing emotional memories [90] or for combining with force feedback to convey pro-social emotions [84].

In addition to tactile driving forms, temperature feedback also plays a role in regulating human emotions. Studies have demonstrated that fear and, to a lesser degree, sadness are associated with "cold" emotions, while joy is perceived as "warm", and anger as a "hot" emotion [9]. Tsetserukou et al. [9] designed the HaptiShiver interface, combining temperature and vibration to convey fear emotions to individuals. However, temperature feedback devices are limited in their capacity for rich pattern expression, and precise control of temperature drivers is challenging, especially for rapid temperature mode switching, often requiring extended periods, especially for transitioning from high to low temperatures.

### 5.3.4 General tactile parameters

The design space of MSTs can be conceptualized as a set of adjustable parameters essential in the development of MSTs. Some of these adjustable parameters are referenced across multiple tactile modalities. While the mechanical properties associated with these parameters vary across different modalities, they yield similar experiential or perceptual outcomes. Thus, we have compiled a range of adjustable parameters mentioned in various tactile interaction techniques, abstracting away from modality-specific mechanical meanings to define these universal parameters (i.e., *Intensity, Pattern, Position, Duration, Sequence, Frequency, Envelope frequency, Amplitude, Speed, Texture, Force direction, Size, Shape*, and *Two-state*)(see Table 1).

These parameters encompass those shared by all modalities, such as Two-state and Speed, as well as specific parameters unique to certain modalities. For example, Texture is a parameter of tactile modality based on material or similar tactile feedback techniques (excluding simulated texture through vibration signals). These parameters might result from the effect of a single mechanical parameter (e.g., Duration, where the mechanical and experiential meanings align) or from the coupling of multiple mechanical parameters (e.g., vibration Intensity, a result of the interaction between frequency and amplitude; Speed, a combination of attack duration and release duration during tactile feedback). These parameters serve as the foundational elements of experience and can be combined to form compound parameters: (i) movement, the overall movement parameter of the tactile interface, formed by combining Position and Sequence; and (ii) dynamic pattern, a combination of multi-actuator frame patterns and sequences or a single actuator's Position and Sequence. Pattern and its enclosed parameter space thus constitute the design space of MST.

**Table 1** General tactile parameters. (Modality: V-vibrotactile, F-Force, M-Mid-air, T-Texture)

| Parameter | Definition | Other Description | Modality |
|---|---|---|---|
| Pattern | Set of positions of actuators active simultaneously. *Requires multiple actuators in MST | *Actuators in positions equal to finger locations [64] *Numbers of contact points [12, 63, 65, 85] *Workplace [79] *Touch range [82] *Operating locations/areas (fixed overall position) [93, 95] | V, F, M, T |
| Position | Overall position of MST. *Remains constant if MST does not move during tactile interaction. | *moved/moving/movements [66, 83, 84] | V, F, M, T |
| Duration | Time during which tactile feedback (a tactile signal) remains non-zero. | *Switch on and off at different speeds (on/off duration) [66] *Speed levels (operation sequence duration) [100] *Poke length [87] *Operating time [93] | V, F, M, T |
| Sequence | Sequence composed of intermittent tactile signals, describing their order, not their duration. | *Gestures with speed levels (intervals exist) [100] *Touch interval [82] *Number of pokes [87] | V, F, M, T |
| Frequency | Excitation frequency of tactile signals, often affecting feedback texture. | *Carrier frequency [94] | V, F, M, T |
| Envelope frequency | Frequency of behavior bundles constituted by a group of tactile feedback actions, typically perceptible in granularity. | ／ | V, F, M, T |
| Amplitude | Magnitude of a tactile signal. *May influence Intensity but does not determine it completely. | *Operation Voltage [93] *Denote amplitude [94] | V, F, M |
| Intensity | Perceived strength or saliency of tactile feedback. | *Force equal to the input [75] *Air jet flow rate [83] *Amplitude of a poke (not driven amplitude) [87] *Pressure [88] | V, F, M |
| Speed | Rate of appearance and disappearance of a tactile signal. *Independent of tactile signal duration. | *Speed of touching/ retreating [82] | V, F, M, T |
| Texture | Fixed physical features of the interface within tactile feedback. | *Combination of material [90] | T |
| Force direction | Direction in which tactile feedback applies force to the skin, tangential or normal. | *force type [85] | V, F, M, T |
| Size | Contact area of the interaction part in MST with the skin. *Excludes actuator count feature in Pattern. | ／ | F, M, T |
| Two-state | Presence or absence state of tactile feedback. | *The presence or absence of vibration [65] | V, F, M, T |

# 6 Design strategies for MSTs

## 6.1 Workflow

The construction of MSTs often adopts user-centered design thinking [58], participatory design [101], and other design methods. Researchers are increasingly emphasizing the integration of user feedback into the design process. When defining tactile-emotion mapping strategies, particularly the exploratory approach, researchers typically follow an user-centered design process, from obtaining haptic descriptions to validating emotional responses [27, 91]. This process, based on the *Find-Fix-Verify (FFV)* [102] pattern introduced by Bernstein has been applied in various studies, and expanded by *Hapticians* (Haptic experience designer [103]). Do et al. further proposed a *Gen-Rank-Verify* user research structure [104] for designing an intelligent wristband capable of generating squeezing sensations, involving (i) Gen: participants creating their optimal squeezing patterns, (ii) Rank: ranking the patterns, and (iii) Verify: evaluating the suitability of the patterns for a smartwatch's vibrational tactile feedback. By incorporating user-generated design elements and involving groups, researchers can overcome subjective biases and achieve rapid and diverse design enhancements.

User-generated design approach also has several limitations. Firstly, some studies have recruited a small number of designers (less than 20 participants[27, 105]) during the design phase, leading to randomness in results. To mitigate this, researchers should aim to expand the user base. Secondly, the mapping low-dimensional tactile information to high-dimensional emotional information is especially challenging for individuals without semantic coding experience or perceptual backgrounds. To lower the design barrier, researchers need to provide basic references and organize workshops for participants. Lastly, a diverse group of participants with various backgrounds, including perceptual experience designers and regular users, should be recruited to ensure balanced insights and prevent design bias.

The core goal is to make tactile-emotion mapping easy to comprehend. To enhance detailed design strategy guidance, researchers can consider collecting the designers' design intentions and the identifiability methods of classifiers[27]. Qualitative analysis methods like thematic analysis[106] can help summarize latent clues in mapping mechanisms. Shared clues that both users can perceive are used as design references to ensure bidirectional flow of emotional tactile information.

## 6.2 Evaluation methods

**Experimental Settings:** The evaluation of MSTs often involves user experiments to assess system performance. Most studies conducted laboratory-based user tests[94], with limited field test verification. Conducting long-term user tests in real-life scenarios can reveal usability, learnability, and potential problems that might not emerge in lab settings. While most works feature limited user test duration due to lab-based experiments, Park et al.[64] recruited 4 couples for a 5-day experiment involving the CheekTouch system. Observing user logs allowed the analysis of user preferences and patterns of use.

**Table 2** Evaluation methods for MSTs.

| Touch | Affect | Stress and Anxiety | Usability |
|---|---|---|---|
| STQ: Social Touch Questionnaire [107] | SAM: Self-Assessment Manikin [108] | STAI: State-Trait Anxiety Inventory [109] | OME: Observational Measurement of Engagement [110] |
| NFT: Need for Touch Survey [111] | PAD: Pleasure, Arousal and Dominance [112] | SSSQ: Short Stress State Questionnaire [113] | SUS: System Usability Scale [114] |
| | MMSE: Mini Mental State Examination [115] | SSR: Subjective Stress Rating [116] | NASA TLX: Task Load Index [117] |
| | PANAS: Positive and Negative Affect Schedule [118] | PSS: Perceived Stress Scale [119] | |
| | BMIS: Brief Mood Introspection Scale [120] | TAI: Test Anxiety Inventory [121] | |

**Evaluation Scale and Content:** The number of participants recruited for evaluation is generally similar to or more than that for the design (generation) phase[75, 93]. Evaluation content includes user *preference*, system *usability*, tactile *perception*, emotional expression *performance*, *usage* patterns, and other potential *impacts*. Researchers often use physiological signals as indicators, such as blood pressure, skin conductivity, heart rate, and EEG signals, along with subjective evaluations collected via standardized instruments like questionnaires, surveys, and scales. We have combined the usage of the surveyed literature and selected some commonly used evaluation instruments for MSTs based on the summary from MacLean et al. [10], as shown in the Table 2.

## 6.3 Ethical considerations

Ethical considerations assume paramount significance for MSTs. User experiments must be conducted under pertinent ethical review, securing user consent prior to engaging in experiments, and implementing measures to safeguard users' well-being during the experiments. These measures encompass practices such as facility disinfection for physical contact requirements [75], conducting experiments within secure and controlled environments, and ensuring the protection of experimental data. With the potential growth of the digital touch market, a scrupulous assessment of data collection scope and permissions for cloud storage becomes indispensable.

The ethical implications extend beyond the developmental phase. In real-life scenarios, issues might emerge that transcend considerations limited to controlled lab settings. It is imperative to solicit feedback from stakeholders subsequent to the integration of these technologies into daily life. For instance, the LoveBomb project [76] underscores the need to investigate the willingness of strangers to receive random affective messages when such technology is employed in public spaces.

The selection of tactile interaction locations mandates contemplation on whether users perceive an invasion of personal space during social interactions. For instance, when crafting the second iteration of the TapTap prototype, Bonanni et al. [97] scrupulously considered the potential feelings of unease, constriction, or infringement that might be evoked by neck-based interaction. MSTs provoke critical examinations of digital intimacy [122]. An exemplar of this ethical debate is manifested by the Kissenger project [75], which instigated discussions on infidelity within interpersonal relationships due to its support for remote kiss interaction. Instances arise where the utilization of such devices may be construed as acts of unfaithfulness toward partners.

The misuse of social tactile interfaces may also pose ethical risks. When the conveyed affective messages can be personalized, particularly in public settings, reasonable restrictions on the content of tactile messages become pivotal to avert the dissemination of harmful information. The LoveBomb project by Hansson et al. [76] exemplifies this concern as it restricts users from dictating message content.

## 6.4 Cultural diversity

Cultural characteristics can be expressed through different social etiquette, among which the popularity of physical contact in social interactions is a typical representation. In cultures like Arab, Latin American, Jewish, and certain African societies, high levels of physical contact, such as arm embracing, hand-holding, and head touching, are prevalent. These cultures are termed *"High-contact culture"* [123] or *"Contact-oriented cultures"* [124]. Conversely, countries like the UK, China, and Japan exhibit minimal physical contact among acquaintances in public, classifying as *"Low-contact culture"* [124]. Even within High-contact or Low-contact cultures, there are variations in touch customs among regions and nations. Fundamental cultural disparities leading to *"culture shock"* necessitate researchers to consider the cultural backgrounds of their target audience in MST studies. This consideration can be approached from the following perspectives:

**Diverse understandings of the same touch behavior:** Cultural backgrounds lead to varying interpretations of the same physical touch behavior. For instance, same-sex physical contact is acceptable and common among friends in China, but in Anglo-American cultures, it might be misconstrued as behavior between romantic partners [125].

**Diverse touch behaviors for the same social interaction:** Within similar social contexts, individuals from different cultural backgrounds may choose distinct physical touch behaviors, even within the same social relationships. This fact is pivotal when researchers select a mediating model for MST. For example, in English-speaking countries, intimate relations between males and females may involve hugging or cheek kissing, while in public settings, Chinese culture lacks habits of kissing or embracing. The most intimate gestures in Chinese culture involve holding hands, linking arms, or a tight shoulder embrace, indicating affection [125, 126]. Furthermore, differences in spatial awareness and social distancing norms [127, 128] lead to varied attitudes among individuals from different cultural backgrounds regarding the acceptability of touch in public social interactions [129].

**Various presentation of the same touch behavior:** The frequency, timing, duration, and intensity of similar social touch behaviors vary across cultures, influencing design specifics of tactile feedback in MSTs such as physical parameters and form factors. Different cultural groups exhibit distinct physical touch behaviors in public conversations. For instance, the handshake—a gesture of courtesy in English-speaking countries—entails a brief handclasp followed by an immediate release, resulting in an increased physical distance. In contrast, Chinese etiquette involves a handshake followed by a closer physical proximity, sometimes with continued handholding or even mutual grip, considered taboo in English-speaking cultures [123].

### 6.5 Future explorations

**Hybrid modality integration for enhanced expressiveness:** An avenue of paramount importance lies in the optimization of MST's expressiveness through the fusion of tactile modalities. Presently, MST's capacity to convey a diverse spectrum of emotions remains a challenge, in part due to the limitations inherent in single tactile modality interactions. These interactions often fall short of the authenticity inherent in real-life scenarios, where multifaceted tactile engagements prevail. Though a select few studies have ventured into the combination of two tactile modalities [44, 84, 88, 92], frequently involving the integration of vibration and pneumatic force feedback, a thorough investigation into refining the spatiotemporal synchrony within hybrid MST is warranted. Such investigations hold the promise of bestowing users with tactile experiences that are not only richer but also multifarious in emotional dimensions.

**Establishing universal tactile-emotion mapping models:** An imperative forward step involves the establishment of a unified framework for tactile-emotion mapping within the context of MST. Presently, the absence of a standardized model for mapping tactile sensations to emotional states impedes seamless comprehension across diverse cultural contexts. This challenge persists even when considering analogous social scenarios and emotional triggers. To bridge this gap, researchers can harness innovative methodologies such as crowdsourced editing. By accumulating a diverse array of mapping design outcomes, researchers can unravel overarching principles. The subsequent analysis of factors contributing to differentiation in mapping outcomes, including demographic nuances, will pave the way for more effective and adaptable tactile-emotion mappings.

**Diversification of emotional expression space:** Expanding the horizons of MSTs' emotional repertoire is an integral pursuit for the future. Currently, MST predominantly captures high-arousal, high-valence emotions, leaving untapped potential within the realm of infrequently experienced emotional states. The holistic development of MST must entail a deliberate diversification strategy, accommodating subtle emotional cues and nuances. This diversification inherently requires the formulation of rigorous assessment methodologies tailored to capture the nuanced spectra of emotions. While subjective evaluation remains pivotal, the integration of pertinent physiological data can bolster the depth and rigor of these assessments.

**Temporal and spatial versatility of MSTs:** Expanding MST's footprint to encompass temporally and spatially disjointed social contexts presents an exciting avenue. Much like contemporary social media platforms transcending temporal and spatial barriers to facilitate visual, auditory, and textual interactions, MST can potentially metamorphose into an "emotional registry" within social landscapes. Noteworthy examples include the Cheektouch system designed by Park et al. [64], which empowers individuals to archive and replay emotional tactile signals, affording them a tangible means to capture, relive, and share emotional connections.

**Human-robot interaction and emotional enrichment:** In summary, the future of MST beckons the refinement of expressive capabilities, the establishment of universal mapping models, the diversification of emotional expression, adaptation to temporally and spatially dislocated contexts, and the fostering of enriched human-robot interactions. These avenues collectively forge a path toward a more nuanced, versatile, and emotionally resonant landscape within the realm of mediated social touch.

**Encouraging soma-based interactive design practices:** In tactile interactions, bodily perceptions significantly influence the effectiveness of emotional communication. This study aims to review the role of bodily locations in MSTs to encourage designers to utilize the Soma-based design approach, leveraging subjective design capabilities from a first-person perspective [130]. This approach explores how interaction interfaces interact with human bodily perception and behavior, aiming to create more humane and natural user experiences. Furthermore, designers can gain a deeper understanding of human sensory and movement patterns, thereby aligning MSTs designs more closely with human physiological and psychological needs [131].

# 7 Conclusion

In our study, we introduced a human-oriented approach to categorize MSTs to understand their emotional expression space, corresponding design space and strategies. Our analysis of the expression space highlighted an arrow-shaped distribution in the two-dimensional VA model, indicating MST's ability to express both Pleasant-Activated and Unpleasant-Deactivated emotions. Moreover, emotions with high arousal were more prevalent in the middle pleasant region, and those with high valence were more expressed in the middle arousal domain. Regarding the design space, we focused on three primary design factors, observing their impact on emotional expression in MSTs. MST's emotional expression capability varies across different body locations, with the forearm and hand emerging as prominent interaction sites. Wearable devices were the most common form factor, followed by desktop devices, with force feedback and vibration being the most utilized tactile modalities.

Based on the above findings, we considered the design elements of MSTs as independent variables and emotional expression space as dependent variables, aiming to understand their relationship. Future research should focus on expanding MSTs' emotional expression scope and refining the tactile-emotion mapping mechanism. Both MST design factors and the demands of emotional communication in social environments influence the emotional expression space. It is necessary to explore MST application scenarios to develop precise design requirements for specific contexts, fostering more tailored MST applications.

*Funding Information*

This work is supported by National Social Science Fund of China (23BZW014).

*Conflict of Interest Statement*

On behalf of all authors, the corresponding author states that there is no conflict of interest.

# A Appendix

## A.1 MST cases

Table 3: MST cases. (Modality: V-vibrotactile, F-Force, M-Mid-air, T-Texture. Form factor: H-Hand held, D-Desktop, N-non, W-Wearable.)

| Body Location | Year | Author | Source | Tactile Modality | Form Factor | Emotion | Ref. |
|---|---|---|---|---|---|---|---|
| Lips | 2012 | Samani et al. | DIS | F | H | Intimacy | [75] |
| Cheek | 2012 | Park et al. | CHI | V | H | Comfort, Whine, grumble | [64] |
| Nape | 2023 | Chew et al. | CHI EA | M | D | Fear, awe | [66] |
| Forearm | 2020 | Nunez et al. | IEEE Haptics Symposium | V F | W D | Pleasantness | [63] |
| Forearm | 2021 | Salvato et al. | IEEE Transactions on Haptics | V | W | Ask for attention, gratitude, happiness, calming, love, sadness | [79] |
| Forearm | 2013 | Huisman et al. | CHI EA | V | W | Anger, fear, sadness, happiness, love, disgust, gratitude, sympathy | [12] |
| Forearm | 2018 | Israr et al. | CHI EA | V | W | Pleasantness | [80] |
| Froearm | 2018 | Culbertson et al. | IEEE Haptics Symposium | V | W | Pleasantness | [65] |
| Froearm | 2020 | Cang et al. | IEEE Haptics Symposium | V | W | Anxious, love, sad, excitement, attention, anger, calm, hey, gratitude, miss | [77] |
| Forearm | 2015 | Knoop et al. | CHI EA | F | W | Pleasantness | [81] |

Table 3: MST cases. (Modality: V-vibrotactile, F-Force, M-Mid-air, T-Texture. Form factor: H-Hand held, D-Desktop, N-non, W-Wearable.)

| Body Location | Year | Author | Source | Tactile Modality | Form Factor | Emotion | Ref. |
|---|---|---|---|---|---|---|---|
| Forearm | 2022 | Zhou et al. | DIS | F | D | Anger, disgust, fear, happiness, sadness, love, gratitude, sympathy | [82] |
| Froearm | 2018 | Tsalamlal et al. | IJHCI | M | D | Pleasantness | [83] |
| Forearm | 2020 | Sato et al. | IEEE RO-MAN | T F | N | Excitement, happiness, contentment, calmness, sleepiness, boredom, sadness, fear, anger | [84] |
| Forearm and hand | 2019 | Kim et al. | CHI | F | N | Love, comfort, happiness, sadness, fear, anger, disgust, surprise | [85] |
| Hand | 2010 | Bickmor et al. | IEEE Transactions on Affective Computing | F | W | Anger, fear, sadness, disgust, happiness, surprise, sympathy, love, pride, envy, embarrassment, ask for attention, gratitude, others | [86] |
| Hand | 2022 | Rognon et al. | Frontiers in Computer Science | F | W | Enthusiastic, happiness, nervousness, serenity, ask for attention, relaxation, stress | [87] |

**Table 3**: MST cases. (Modality: V-vibrotactile, F-Force, M-Mid-air, T-Texture. Form factor: H-Hand held, D-Desktop, N-non, W-Wearable.)

| Body Location | Year | Author | Source | Tactile Modality | Form Factor | Emotion | Ref. |
|---|---|---|---|---|---|---|---|
| Hand | 2022 | Price et al. | ACM Transactions on Computer-Human Interaction | V F | W | Joy, loneliness, distress, relaxation, ecstasy, fear, longing, a sense of peace, anger, comfort, happiness. | [88] |
| Hand | 2016 | Ahmed et al. | ACM International Conference on Multimodal Interaction | V F | W | Fear, anger, happiness, neutrality, sadness | [44] |
| Hand | 2021 | Ju et al. | CHI EA | V | H | Joy, anger, sadness, relaxation | [27] |
| Hand | 2001 | Hansson et al. | CHI EA | V | H | Sorrow | [76] |
| Hand | 2008 | Eichhorn et al. | International conference on Human computer interaction with mobile devices and services | F | H | Intimacy | [74] |
| Hand | 2007 | Smith et al. | IJHCS | V | D | Anger, delight, relaxation, unhappiness | [89] |
| Hand | 2014 | Nakanishi et al. | CHI | F | D | Empathy | [73] |
| Hand | 2022 | zhen et al. | West Leather | T | D | Empathy, satisfaction, joy, encouragement | [90] |
| Hand | 2015 | Obrist et al. | CHI | M | D | Pleasantness | [91] |

**Table 3**: MST cases. (Modality: V-vibrotactile, F-Force, M-Mid-air, T-Texture. Form factor: H-Hand held, D-Desktop, N-non, W-Wearable.)

| Body Location | Year | Author | Source | Tactile Modality | Form Factor | Emotion | Ref. |
|---|---|---|---|---|---|---|---|
| Finger | 2017 | Singhal et al. | ACM conference on computer supported cooperative work and social computing | V | W | Intimacy, empathy | [132] |
| Shoulder blade | 2006 | Bonanni et al. | CHI EA | V | W | Miss | [97] |
| Chest and shoulder | 2012 | Teh et al. | IEEE Haptics Symposium | F | W | Pleasantness | [61] |
| Back and stomach | 2005 | Mueller et al. | CHI EA | F | W | Intimacy | [62] |
| Chest and back | 2009 | Tsetserukou et al. | International conference on affective computing and intelligent interaction and workshops | F | W | Empathy | [9] |
| Stomach | 2009 | Tsetserukou et al. | International conference on affective computing and intelligent interaction and workshops | V | W | Love, joy | [9] |

**Table 3**: MST cases. (Modality: V-vibrotactile, F-Force, M-Mid-air, T-Texture. Form factor: H-Hand held, D-Desktop, N-non, W-Wearable.)

| Body Location | Year | Author | Source | Tactile Modality | Form Factor | Emotion | Ref. |
|---|---|---|---|---|---|---|---|
| Back | 2017 | Neidlinger et al. | TEI | V F | W | Awe | [92] |
| Neck and back | 2015 | Arafsha et al. | Multimedia Tools and Applications | V | W | Love, joy, surprise, anger, sadness, fear | [93] |
| Back | 2021 | Foo et al. | IMWUT | F | W | Sadness, anger, happiness, fear, gratitude, love, calm, attention | [95] |
| Back and thigh | 2018 | Chandra et al. | Ubicomp | V | D | Calming, excitement, annoyance | [94] |
| Back and thigh | 2020 | Chandra et al. | IEEE Haptics Symposium | V | D | Excitement, calmness, sadness, nervousness, neutrality | [96] |
| Foot | 2016 | Turchet et al. | IEEE Transactions on Affective Computing | V | W | Happiness, aggression, tenderness, sadness, neutrality | [98] |